# LLM hallucinations in the wild: Large-scale evidence from non-existent citations

Zhenyue Zhao[1†], Yihe Wang[1,2,3†*], Toby Stuart[4], Mathijs De Vaan[4], Paul Ginsparg[1], Yian Yin[1]

[1]Department of Information Science, Cornell University, Ithaca, NY, USA

[2]Department of Sociology, University of California Los Angeles, Los Angeles, CA, USA

[3]Department of Computer Science and Technology, Tsinghua University, Beijing, China

[4]Haas School of Business, University of California Berkeley, Berkeley, CA, USA

[†]These authors contributed equally to this work.

[*]Research conducted when visiting Cornell and UCLA.

**Large language models (LLMs) are known to generate plausible but false information across a wide range of contexts, yet the real-world magnitude and consequences of this hallucination problem remain poorly understood. Here we leverage a uniquely verifiable object — scientific citations — to audit 111 million references across 2.5 million papers in arXiv, bioRxiv, SSRN, and PubMed Central. We find a sharp rise in non-existent references following widespread LLM adoption, with a conservative estimate of 146,932 hallucinated citations in 2025 alone. These errors are diffusely embedded across many papers but especially pronounced in fields with rapid AI uptake, in manuscripts with linguistic signatures of AI-assisted writing, and among small and early-career author teams. At the same time, hallucinated references disproportionately assign credit to already prominent and male scholars, suggesting that LLM-generated errors may reinforce existing inequities in scientific recognition. Preprint moderation and journal publication processes capture only a fraction of these errors, suggesting that the spread of hallucinated content has outpaced existing safeguards. Together, these findings demonstrate that LLM hallucinations are infiltrating knowledge production at scale, threatening both the reliability and equity of future scientific discovery as human and AI systems draw on the existing literature.**

The propensity of Large Language Models (LLMs) to generate plausible but false information, often referred to as "hallucination", remains an unresolved challenge even in state-of-the-art models [1]. While model developers recommend that users verify output for potential errors, in practice, many hallucinations go unrecognized. When it appears in substantive documents, hallucinated content then becomes incorporated into downstream outputs, possibly eroding the factual basis for high-stakes decisions [2]. Recent reports have documented alarming instances of hallucinated content entering government policy reports [3], court filings [4], [5], and scientific publications [6], [7], [8], spurring debate over regulation and appropriate use of AI tools [9].

This growing concern highlights the need to audit the prevalence of hallucinations "in the wild". While AI labs study hallucinations in simulated or controlled settings, we know less about their prevalence in deployed use. A systematic understanding of real-world hallucinations is important for the technical community that develops foundation models, as well as for institutions that rely on and vet knowledge work, such as courts, the patent office, media outlets, medical bodies, and of course, peer review. At present, we lack such evidence.

Hallucinations are difficult to study at scale because they are often embedded in long, unstructured text. Quantifying their incidence therefore requires extracting large samples of claims, retrieving factual evidence from external sources, and assessing the validity of each claim against a ground truth. Recent attempts to measure hallucinations have relied on human evaluations that are costly to scale [10], or automated LLM-based pipelines that might reintroduce model-specific biases into the analysis [6], [7].

We present a large-scale audit of hallucinated content in the scientific literature, the most measurable instance of a phenomenon that likely extends across all knowledge work. While the widespread use of GenAI in scientific production has already reshaped productivity dynamics and quality signals [11], the cumulative nature of scientific research [12] makes the knowledge ecosystem vulnerable to perpetuating errors [13], [14]. As researchers increasingly use LLMs to source citations, they risk introducing  hallucinated artifacts into the permanent record [15].

Science provides a compelling laboratory because citations offer a traceable, and therefore tractable, ground truth for studying hallucinations. Unlike most corpora, the existence of scientific citations is unambiguous. A cited reference exists or it does not. This makes the scientific literature an ideal case study for developing a scalable hallucination detection methodology. Science is also arguably the best-case domain for detecting and removing hallucinated information because it has a rich indexing system, strong citation norms, and layered editorial review. This suggests that the prevalence we measure here likely understates what is occurring in less structured knowledge domains.

Scholars have proposed several approaches to studying hallucinations in science. First, experiments prompting LLMs to suggest references have documented hallucinated citations [16], [17], [18], [19]. Although estimates vary substantially across fields, models, and specific prompts, these magnitudes likely represent an upper bound because they test model behavior in isolation, without accounting for error detection from human oversight. Second, researchers have developed tools to identify potentially hallucinated citations and applied them to selected domain-specific corpora [6], [7], [10] and preprint servers [20]. These studies provide early evidence of hallucinated citations, often at the scale of tens to hundreds of suspected cases. At the same time, scientific references span vast sources (e.g., non-English journals and media outlets), creating a "false positive" challenge that often requires manual validation [15]. Moreover, most existing studies focus on a narrow set of venues with especially high AI use, suggesting that findings may not generalize to the broader scientific literature.

We take a different approach. Rather than flagging individual hallucinations, we track the rate and characteristics of unmatched citations over time, comparing post-LLM trends against pre-LLM baselines. Because unmatched citations existed before LLMs, our design does not treat every unmatched citation as hallucinated. Instead, we use pre-LLM unmatched citations as a baseline for ordinary bibliographic and matching errors, and interpret the post-LLM excess of unmatched citations as the population-level signature of hallucination. This design yields an estimate of hallucination prevalence net of baseline error, while also allowing us to examine demographic and structural correlates of hallucinated citations. Applying this method to four large corpora, we conduct a population-scale audit of citation hallucinations to provide evidence of how LLM-generated errors have entered the scientific record.

**Estimating hallucinated references at scale**

We first develop a pipeline to verify the existence of academic references. We extract references from raw manuscript files, then parse structured fields from each reference string. Following existing practices in citation hallucination detection [21], we focus on one specific type of hallucinated reference: those with non-existent titles. For all references, we query a locally hosted Elasticsearch index constructed from Semantic Scholar and OpenAlex. Each parsed title is used to retrieve candidate matches, and we determine whether any candidate constitutes a valid match using string similarity-based criteria. This approach matches 95.1% of references.

By manually validating many of the unmatched cases, we find that false negatives most often arise for two reasons: the title corresponds to a non-academic source, or it was imperfectly parsed. To address these cases, we route all unmatched entries through an LLM-based cleaning step. Using GPT-4o-mini, we identify and exclude reference strings that are unlikely to be academic references, reducing the unmatched rate to 2.33%. We then refine title extraction and feed these updated titles back into the Elasticsearch pipeline, further reducing the unmatched rate to 1.54%.

Finally, to capture cases that may fall outside our databases, we perform an additional lookup using a third-party API to query Google Scholar, one of the most comprehensive citation repositories. Based on the search results, we treat a reference as valid if Google Scholar returns either a clearly matching record or consistent evidence that the same-titled work has been cited across multiple papers. References that remain unverifiable after this step constitute our final pool of "unmatched" citations. More details on the detection methods, thresholds, and manual validations are available in Supplementary Information S2.1, S3.1 and S4.

To quantify the dynamics of hallucinated citations at scale, we collect and examine four bibliometric datasets (SI S1, S3.4): (1) arXiv, covering 1,465,145 preprints across mathematical, physical, and computational sciences (Jan 2020–Aug 2025). We extract references directly from the source bibliography files for the 60.9% of submissions in LaTeX format and apply GROBID to recover citations from raw PDFs for the remainder (44,107,529 citations). (2) bioRxiv (261,928 preprints), which spans a broad range of biological and life science fields. We collect 21,183,111 XML-formatted citation records generated by the platform's in-house processing algorithm. (3) Social Science Research Network (SSRN, 421,698 preprints), a working paper repository hosting manuscripts in the social sciences, law and the humanities. While full texts are not publicly available from SSRN, we retrieve metadata totaling 26,815,043 citations via Crossref for manuscripts assigned an SSRN DOI. To test whether hallucination patterns are confined to non–peer-reviewed preprints, we extend our analysis to (4) PubMed Central (PMC), a leading corpus of peer-reviewed, full-text journal publications. For computational efficiency, we focus on a 10% random sample of recent papers (374,807 manuscripts between 2020 and 2025), with 19,245,787 references extracted. Together, these datasets are among the largest within their respective domains, yet they systematically differ in scientific fields, publication formats, and citation-parsing techniques. (SI S1).

Figure S2 plots the fraction of cited references that cannot be matched to the "ground-truth" citation databases over time. Across all datasets, the unmatched rate remains relatively stable through late 2022, corresponding to the baseline error rate of our verification pipeline. Manual inspection suggests that these unmatched citations stem from edge cases, such as formatting artifacts or items not indexed by mainstream citation databases. The unmatched rate then increased sharply in 2023, likely representing the emergence of hallucinated citations in science. Throughout the rest of this paper, we use "hallucinated citations" to refer to the estimated excess of unmatched citations above the pre-LLM baseline, rather than classifications of individual references.

## Results

### The rise and distribution of hallucinated citations

We estimate the dynamics of hallucinated citations using a regression framework (S2.2). Figure 1 plots the estimated coefficients, highlighting a continuous growth of hallucinated citations, reaching levels of 0.39% (arXiv), 0.21% (bioRxiv), 1.91% (SSRN), and 0.27% (PMC) as of August 2025. The steepest rise begins in mid-2024, roughly 18 months after the initial release of ChatGPT in late 2022. Unlike the rapid uptake of LLMs for writing assistance documented in early 2023 [22], [23], the delayed inflection may reflect the evolution of LLM functionality to include AI search and agentic research assistants that automate the generation of citations from live web content.

We further estimate the magnitude of hallucinated citations by calculating the observed excess number of unmatched references (SI S2.2). Our estimates suggest a substantial volume of hallucinated citations, reaching a monthly estimate of 3,353 (arXiv), 478 (bioRxiv), 767 (SSRN), and 8,140 (PMC) in August 2025. Assuming these trends persisted through year end, these four corpora, which cover only a fraction of the scientific literature, would include 146,932 hallucinated citations in 2025 alone. SI S3.1 documents additional sensitivity tests and cross validations of our estimates, showing highly consistent results. Given that our verification method is lenient with respect to formatting or spelling irregularities and focuses purely on the factual validity of reference titles, these estimates are very likely to be a lower bound on the true prevalence of AI-hallucinated citations in science.

The surge documented in Fig. 1a-d mirrors recent, anecdotal reports of generative AI "slop". A natural hypothesis is that hallucinated citations arise from a small number of actors producing deeply contaminated manuscripts. To test this, we examine the distribution of the ratio of unmatched to accurate references in each paper. Figure 1e-h rejects the "most-or-nothing" pattern at the paper level. Instead of a few manuscripts with a high fraction of hallucinated citations, many papers contain a modest proportion of hallucinated references. We observe a growing proportion of papers containing at least one hallucinated reference, with increases ranging from 2 to 23 percentage points (pp) across datasets. Indeed, the fraction of papers with 50% or more unmatched references has increased only slightly over the observation window (arXiv: 0.102 pp; bioRxiv: 0.172 pp; SSRN: 2.016 pp; PMC: 0.018 pp), at rates far slower than for papers with 0–10% unmatched references (arXiv: 3.615 pp; bioRxiv: 1.462 pp; SSRN: 7.373 pp; PMC: 1.430 pp). This distribution refutes the "few bad apples" narrative, and suggests more widespread contamination, where non-existent sources are sparsely embedded within (seemingly) legitimate papers.

A plausible explanation for these trends is that LLMs generate mixtures of accurate and hallucinated citations, and authors adopt some of these suggestions without fully vetting them. Experimental studies from 2023 to 2025 have found that the rate of hallucination in LLM-generated citations ranged from 19.9% to 91.4%, varying widely across models and time [16], [17], [18], [19], [24]. Breaking down our analysis by field, we find widespread citation hallucinations (Fig. 1i-j), with markedly higher rates in the social sciences and computer science (SI S2.3). These fields coincide with areas where LLM-assisted writing is reported to be particularly prevalent.

To further examine the link between non-existent citations and LLM use, we apply a standard approach that estimates the intensity of LLM use in scientific writing (SI S2.4) to all paper abstracts in the arXiv corpus. Fig. 1k confirms a strong association between estimated LLM use in writing and hallucination rates at the subfield level, where fields with higher inferred LLM usage exhibit significantly higher rates of hallucinated citations (Pearson correlation $r = 0.441$, $P < 0.001$). This relationship holds at the paper level. Manuscripts with higher estimated LLM usage contain a higher rate of hallucinated citations (Fig. 1l). These results are consistent with LLM use being a major driver of the rise in hallucinated citations.

**Hallucination citers and their beneficiaries**

We next ask how hallucinated citations intersect with the structure of scientific production (Fig. 2). We compare papers containing unmatched citations to a set of control papers matched on publication time and scientific field. To isolate the distinct properties of LLM-induced hallucinations, we model unmatched citations as a contaminated mixture of baseline bibliographic errors and AI-generated fabrications, assigning each citation $i$ (cited in time $t$) two probabilities: $p_t$, the probability that the citation is LLM-hallucinated, and $q_t$, the probability that the citation is a conventional match failure of non-LLM origin. Matched references take $p = q = 0$ by definition, while for unmatched references we have $p + q = 1$. We then estimate the following regression model:

$$y_i = \alpha + \gamma_{\text{category}} + \delta_t + \beta p_t + \theta q_t + \varepsilon_i$$

where $\beta$ captures the deviations of hallucinated citations (relative to matched controls, see SI S2.5 for more details).

Prior work on technology adoption suggests that younger workers are more likely to adopt new productivity tools, including LLMs [22], [25]. Less experienced authors also have less familiarity with existing literature, making it more costly for them to identify non-existent citations. To examine this, we count the number of prior papers (up to 2022) produced by an indexed manuscript's last author (SI S2.6). Fig. 2a confirms that across all datasets, authors citing hallucinated sources (hereafter, "hallucination citers") have substantially fewer publications than members of a matched control group: 62.1% fewer in arXiv, 62.6% in bioRxiv, 73.2% in SSRN, and 27.4% in PMC. Hallucination citers also are much more likely to have *zero* prior publications (32.1% in arXiv, 9.3% in bioRxiv, 51.0% in SSRN, and 3.7% in PMC). Robustness checks using alternative authorship positions, productivity measures, and an alternative control group yield similar patterns (SI S3.3).

Strikingly, this productivity gap between hallucination citers and matched controls has closed by 2025, indicating a marked increase in scientific output among hallucination citers. Comparing publication counts before 2023 to those in 2025, Figure 2a shows the relative productivity increase among hallucination citers to be 1.93× in arXiv, 2.22× in bioRxiv, 3.13× in SSRN, and 1.33× in PMC, when compared to matched, non-hallucination-citing authors. Further, hallucination rates attenuate sharply with team size (Fig. 2b). These results appear consistent with recent evidence that LLM use boosts scientific output [11], [26], especially for less-experienced researchers [27] who also may have fewer resources at their disposal. This means that the authors most likely to introduce hallucinations into the scholarly record now have a high rate of manuscript production. By lowering barriers to entry, LLMs may equip a fringe of inexperienced or sloppy researchers to produce a high volume of inaccurate works, amplifying the spread of unreliable information in science.

We next ask: do hallucinated citations assign credit to actual (existing) scientists, and if so, what are the characteristics of these recipients? To examine this, we extract author names from arXiv citations and link these to existing author profiles. To disambiguate same-name authors, we construct text-based, paper-level embeddings [28] and restrict attributions to cases where the citing paper falls close to the cited author's existing body of work (SI S2.8, S3.5, Fig. S4). As we would expect, we find that authors listed in hallucinated citations are 22.2% less likely to match any existing publication profile (Fig. 2c). This means that hallucinated citation titles frequently co-occur with invented author names.

Examining hallucinated paper titles that match to actual author profiles, hallucinated citations disproportionately credit authors with high prior productivity (68.8% more publications) and high citation impact (58.3% more citations), compared to a matched sample of real citations from the same month and field. Further, we find suggestive evidence of reinforced gender disparity (SI S2.8), where hallucinated citations disproportionately credit authors with male names (6.4 percentage points, corresponding to a 7.6% relative increase), echoing recent findings on sociodemographic distortions in LLM outputs [29], [30].

At the team level, hallucinated references also appear to deviate from conventional authorship patterns (SI S2.8). Despite the increasing dominance of team science [31], hallucinated references tend to credit smaller teams. They also deviate from established author-ordering conventions. In many scientific fields, the last author is typically a more senior and established investigator than the first author. We quantify this as the fraction of references where the last author has greater scientific productivity or impact than the first. In hallucinated references, this first–last author hierarchy is 12 percentage points weaker. Together, these patterns suggest that LLM-generated references deviate from the typical social organization of scientific authorship.

As LLMs are now routinely used as literature discovery tools, a critical question is whether some of these biases extend to *valid* LLM-generated citations. Since papers containing hallucinations (Fig. 1e–h) probably incorporate other LLM-generated references, we ask whether valid citations in these papers also skew toward prominent authors. We find suggestive support for this prediction. Comparing only the valid references in papers containing hallucinations to the reference lists of control papers precisely matched on year, month, field, subfield, and reference list length, citations in hallucination-containing papers indeed skew toward more productive and impactful authors. These results indicate that visible hallucinations are likely just the tip of the iceberg: the broader reallocation of scientific credit likely operates through the much larger set of real LLM-recommended citations, which may largely follow the same preference for already-prominent scholars.

**Systemic responses and failure modes**

The results establish that hallucinated citations are widespread, disproportionately produced by junior researchers, and they channel credit toward established scholars. Next, we ask, to what extent do existing quality-control mechanisms detect and correct them?

We first examine the moderation process on arXiv, a major online preprint platform, by analyzing a unique dataset of 30,981 manuscripts rejected by moderators (assisted by automated screening tools; see SI S2.9 for details). Compared with Fig. 1a, rejected manuscripts have a substantially higher hallucination rate than accepted ones, reaching 2.2% by August 2025 (4.5 times the rate observed among accepted manuscripts). While arXiv moderation is not designed to systematically audit reference lists, it nevertheless rejects a disproportionate share of manuscripts with hallucinated references, consistent with the view that hallucinated citations co-occur with other quality concerns (Fig. 3a inset). Yet at the same time, the rapid growth in submissions and in hallucinated references has far outpaced the current safeguard: we estimate that 78.8% of non-existent citations pass moderation and appear on the platform (Fig. 3a). Amidst the rapid increase in rejection rates since late 2022 [32], manuscripts that are screened disproportionately contain hallucinated citations, but the majority of hallucinated content still enters the scholarly record.

When hallucinated citations survive initial screening on the preprint archives, editorial and peer review might still catch them before publication. To test this, we trace 2,241 bioRxiv preprints containing unmatched references to their published versions in PubMed Central. Among the hallucinations present in preprints, 85.3% of them persist into the published record (SI S2.7), suggesting that current journal practices also do not reliably contain the problem.

Are hallucinated citations confined to lower-tier journals [33]? Examining biomedical papers in PubMed Central, we stratify journals into deciles by their scientific impact (journal-level average hit rate, see SI S2.10 for more details). Consistent with our expectation, journals in the lowest and highest deciles exhibit markedly higher and lower rates of hallucinations, respectively (Fig. 3b). However, there is no consistent gradient between the two extremes, indicating that most journals — including many relatively high-impact ones — remain susceptible, with hallucinated citations appearing across the full impact spectrum.

Once published, hallucinated citations diffuse through the bibliographic sources that researchers and AI systems treat as factual. When authors or automated tools attempt to verify a reference, they consult scholarly search engines and citation databases. As non-existent references surface in these systems as standalone bibliographic entries, in addition to appearing in the full text of citing papers, they become part of the permanent record. Cross-validating unmatched references against Google Scholar (SI S2.11), we do indeed find a growing number of entries that cannot be linked to any real publication yet already appear as references in other papers (Fig. 3c). The pattern replicates across all four corpora (Fig. S10). The accumulation of hallucinated citations within bibliometric databases may begin to erode the mechanisms we have to detect them.

## Discussion

This work provides the first large-scale, cross-disciplinary audit of LLM hallucinations in real-world knowledge work. By developing a reference verification pipeline and applying it to arXiv, bioRxiv, SSRN, and PubMed Central, we document a sharp rise in hallucinated citations beginning in mid-2024, conservatively estimated at more than 146,932 hallucinated references in 2025 alone. Notably, despite concurrent advances in reasoning models and retrieval-augmented generation systems, citation hallucinations continue to rise through our data cutoff in late 2025 with no signs of plateauing.

Non-existent citations offer a distinctive measurement advantage for studying hallucinations in text. They are a class of error that should, in principle, occur with near-zero probability in carefully produced scholarly work. The existence of a factual ground truth makes them a rare commodity in a landscape where most hallucination detection relies on costly human evaluation or LLM-based classifiers that risk reintroducing the biases under study. By operating at the corpus level rather than attempting to classify individual documents, our approach allows us to conservatively estimate prevalence while simultaneously decomposing the structural features of the phenomenon: who produces hallucinated citations, who benefits from them, and how they propagate through the knowledge system. The prevailing discourse in science has framed hallucinated citations as a problem of individual carelessness, amenable to detection and regulation at the manuscript level. Our findings point to a different, systems-level concern. Contamination of the literature is diffusely distributed, but it is disproportionately produced by researchers for whom LLMs have sharply lowered the barriers to scientific production. Their rate of manuscript production has also increased sharply. The result is a low-grade, widespread contamination whose biases channel credit toward already prominent and predominantly male scholars, and whose effects likely extend well beyond hallucinated citations into the broader body of LLM-influenced references.

Once in the published record, hallucinated citations may compound through two channels. First, citation propagation is path-dependent: authors and automated tools routinely build on predecessor reference lists, and our Google Scholar analysis suggests that hallucinated entries already are becoming embedded in the bibliographic infrastructure that researchers treat as ground truth. Second, LLMs are frequently trained on the same open-access corpora we examine, creating a feedback loop in which tomorrow's models will learn today's hallucinations [34].

Our analysis has limitations. The pipeline may produce false negatives for niche venues and heavily mathematical documents, and it has limited coverage for fields whose citation norms omit titles. False positives may arise when LLMs generate a real title with incorrect metadata, though existing estimates suggest such cases are relatively minor compared with the magnitudes we report [20]. More fundamentally, hallucinated titles represent the most detectable form of the problem. The more prevalent and harder-to-detect variant—real citations deployed to support claims the cited references do not actually make [35], [36], [37], [38]—remains an open challenge for which reliable detection methods remain under active development.

A growing ecosystem of agentic verification tools—automated reference validators [39], observability platforms, and evaluation frameworks—offers cautious optimism for the specific class of errors we study. In a domain where citations are structured, indexed, and queryable, these tools may substantially reduce hallucination rates if publishers and institutions adopt them. But citation verification is among the easiest hallucination detection problems: the objects are discrete, the ground truth is well-defined for now, and the lookup infrastructure already exists. The harder problem, detecting claims that misrepresent the content of real sources, or hallucinated assertions embedded in unstructured prose, remains far less tractable, and the gap between what verifiers can detect and what LLMs can produce is wide and probably widening.

This asymmetry is what makes our findings generalizable. Science is arguably the best-case domain for hallucination detection: it has rich indexing systems, large-scale bibliographic databases, strong norms of citation, layered editorial review, and now a rapidly developing toolkit of automated verifiers. Yet even here, hallucinated content enters the record, persists from preprint through publication, and compounds through path-dependent citation networks and model training loops. In domains that lack comparable verification infrastructure, such as government reports, legal filings, clinical documentation, corporate knowledge bases, and journalism, the same class of errors is likely higher in prevalence, harder to detect, and less likely to be corrected. These domains also lack the structured objects that make agentic verification feasible in science; there is no equivalent of a comparable citation index for a factual claim in a policy memo or a clinical note. The problems we document therefore extend beyond science. Science is the most measurable instance of what is likely a far broader phenomenon: the infiltration of AI-hallucinated content into the knowledge systems on which modern institutions make consequential decisions. The challenge ahead is to detect and remove diffuse hallucinations from the vast landscape of unstructured knowledge work.

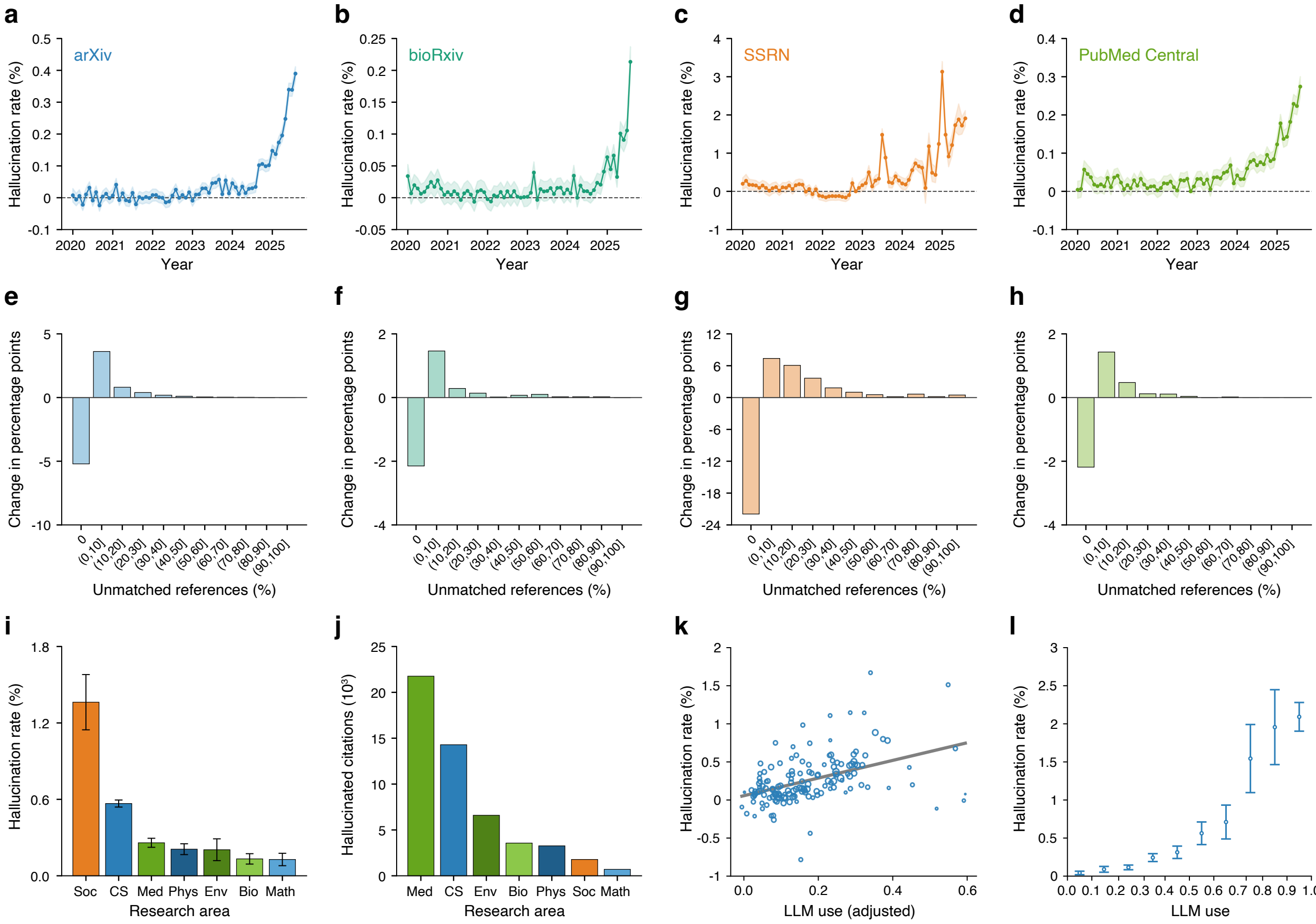

**Fig. 1. The rise and distribution of hallucinated references. a-d**, Quantifying citation hallucination rates through a regression-based approach. All four corpora show a sharp increase after widespread LLM adoption, with the steepest growth beginning in 2024. **e-h**, Changes in the paper-level distribution of unmatched-reference shares. The increase is concentrated among papers containing a small fraction of such references, rather than a small number of heavily contaminated manuscripts, indicating diffuse contamination across the literature. **i-j**, Field-level heterogeneity in citation hallucinations. We measure the percentage (**i**, as of 2025) and total number (**j**, as of 2025) of hallucinated citations by scientific fields, with social science ("Soc") papers estimated from SSRN data, computer science ("CS"), mathematics ("Math"), and physics ("Phys") papers estimated from arXiv data, and medicine ("Med"), biology ("Bio"), and environmental science ("Env") papers estimated from PMC data. **k**, Across arXiv subfields, estimated hallucination rates correlate with estimated LLM use in scientific writing ($r$ = 0.441, $P$<0.001). **l**, At the paper-level, manuscripts with stronger linguistic signatures of AI-assisted writing contain higher rates of hallucinated references.

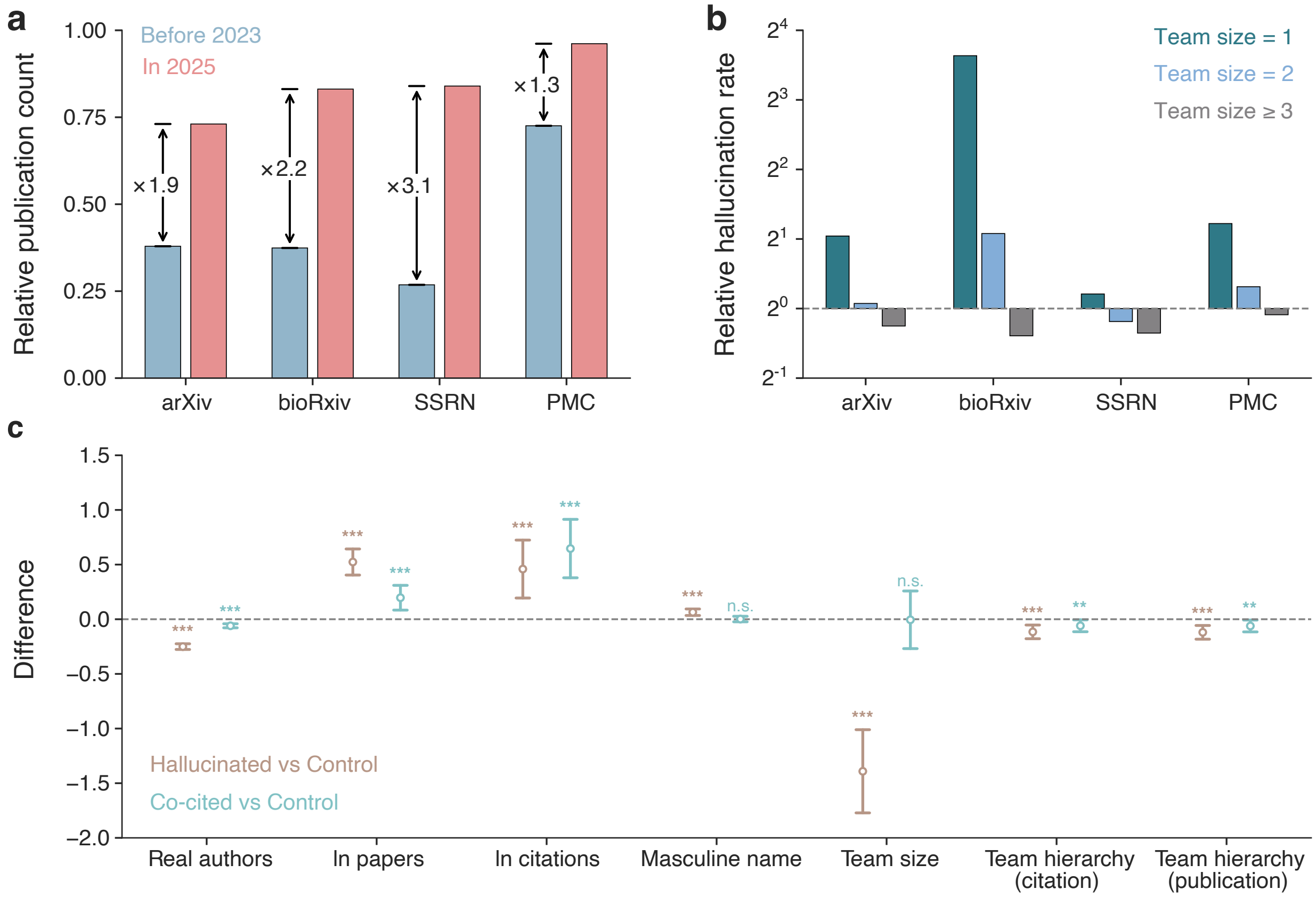

**Fig. 2. Hallucination citers and the unequal allocation of scientific credit. a,** Hallucination citers had substantially fewer prior publications before widespread LLM adoption (pre-2023), relative to the matched controls. However, the gap has largely closed in 2025. **b,** Estimated hallucination rates by team size, normalized by the overall hallucination rate in each corpus. Solo-authored and small-team papers have substantially higher hallucination rates than larger-team papers. **c,** Hallucinated citations are less likely to map to real author profiles. Yet conditional on mapping, however, they disproportionately credit more productive, more highly cited, and male-name authors. Teams cited in hallucinated citations also deviate from conventional authorship patterns, including smaller team sizes and weaker first-last author hierarchy. These biases largely extend to valid references co-cited with hallucinated references. ***$P < 0.01$, **$P < 0.05$, *$P < 0.1$; Error bars represent the standard error of the mean.

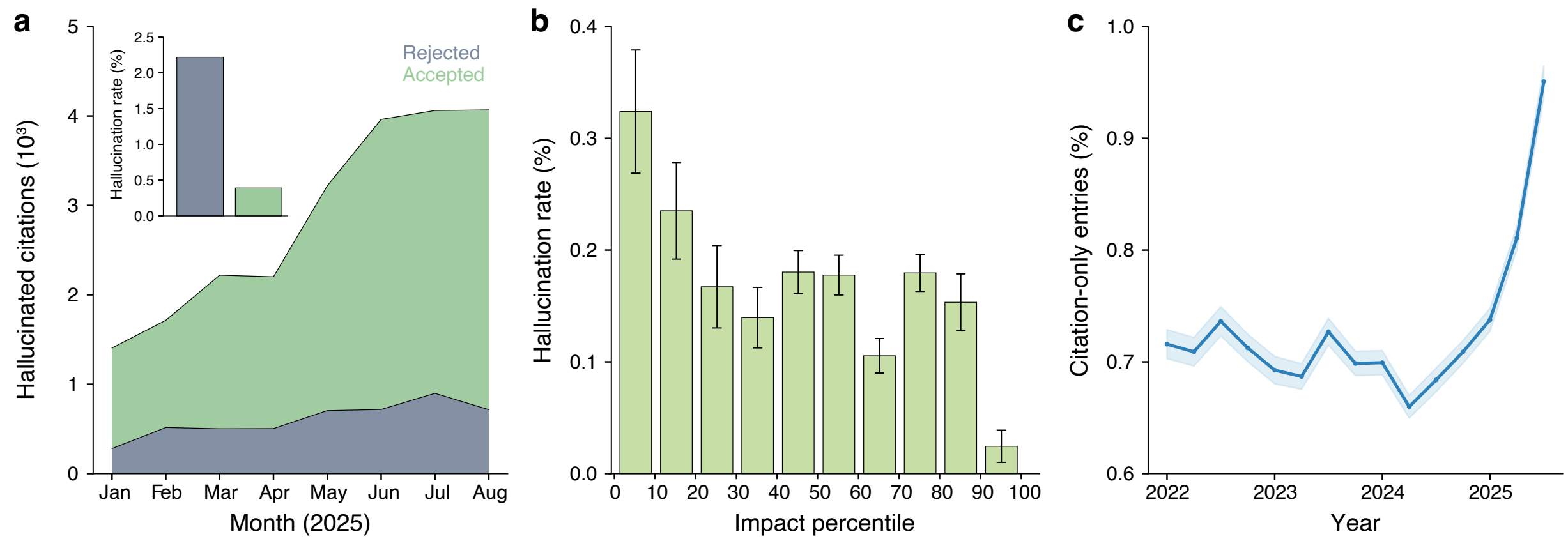

**Fig. 3. Existing safeguards detect some hallucinated references but leave substantial leakage.** **a**, Hallucination rates are higher in arXiv submissions rejected during moderation (inset). Yet most non-existent references still pass moderation and appear on the platform. **b**, Hallucination rates in PubMed Central papers by journal impact percentile. Journals in the lowest and highest deciles exhibit markedly higher and lower hallucination rates, respectively. Yet across the broad middle range, hallucination rates fluctuated between 0.11% and 0.24%, with no clear monotonic association with journal impact. **c**, Growth of citation-only entries. We measure the rate of references that cannot be mapped to real publications but already appear as references in Google Scholar. The rate has increased steeply since 2024, a pattern that replicates across all datasets (Fig. S10).